\documentclass[conference,a4paper]{IEEEtran}
\IEEEoverridecommandlockouts

\usepackage[numbers]{natbib}
\usepackage[cmex10]{amsmath}
\usepackage{amssymb,amsfonts}
\usepackage{dblfloatfix}
\usepackage{orcidlink}
\usepackage{hyperref}
\hypersetup{
    colorlinks=true,
    citecolor=blue,
    linkcolor=blue,
    urlcolor=blue
}

\usepackage{booktabs}       % For \toprule, \midrule, etc.
\usepackage{threeparttable} % For the notes and captions
\usepackage{xcolor}         % For \textcolor
\usepackage{siunitx}        % For S columns
\usepackage{graphicx}       % Standard requirement

\usepackage{graphicx}
\graphicspath{{Figures/PDF/}{Figures/PNG/}}

\usepackage{booktabs}
\usepackage{siunitx}
\usepackage{tabularx}
\usepackage{makecell}
\usepackage{enumitem}
\usepackage{threeparttable}
\usepackage{xcolor}
\usepackage{booktabs}

\usepackage{eso-pic}

\author{
\IEEEauthorblockN{
\begin{minipage}{\textwidth}
\centering
\begin{tabular*}{\textwidth}{@{\extracolsep{\fill}}cccc}
\begin{tabular}{c}
Nichula Wasalathilaka\orcidlink{0009-0003-5662-009X}\\
\textit{University of Peradeniya}\\
Peradeniya, Sri Lanka\\
e20425@eng.pdn.ac.lk
\end{tabular}
&
\begin{tabular}{c}
Dineth Perera\orcidlink{0009-0009-1295-2547}\\
\textit{University of Peradeniya}\\
Peradeniya, Sri Lanka\\
e21291@eng.pdn.ac.lk
\end{tabular}
&
\begin{tabular}{c}
Oshadha Samarakoon\orcidlink{0009-0000-9337-0198}\\
\textit{University of Peradeniya}\\
Peradeniya, Sri Lanka\\
e21345@eng.pdn.ac.lk
\end{tabular}
&
\begin{tabular}{c}
Buddhi Wijenayake\orcidlink{0009-0001-2624-0251}\\
\textit{University of Peradeniya}\\
Peradeniya, Sri Lanka\\
e19445@eng.pdn.ac.lk
\end{tabular}
\end{tabular*}
\end{minipage}
}

\vspace{0.4cm}

\IEEEauthorblockN{
\begin{minipage}{0.78\textwidth}
\centering
\begin{tabular*}{\textwidth}{@{\extracolsep{\fill}}ccc}
\begin{tabular}{c}
Roshan Godaliyadda\orcidlink{0000-0002-3495-481X}\\
\textit{University of Peradeniya}\\
Peradeniya, Sri Lanka\\
roshang@eng.pdn.ac.lk
\end{tabular}
&
\begin{tabular}{c}
Vijitha Herath\orcidlink{0000-0002-2094-0716}\\
\textit{University of Peradeniya}\\
Peradeniya, Sri Lanka\\
vijitha@eng.pdn.ac.lk
\end{tabular}
&
\begin{tabular}{c}
Parakrama Ekanayake\orcidlink{0000-0002-5639-8105}\\
\textit{University of Peradeniya}\\
Peradeniya, Sri Lanka\\
mpbe@eng.pdn.ac.lk
\end{tabular}
\end{tabular*}
\end{minipage}
}
\thanks{The authors acknowledges the support received from the LK Domain Registry in publishing this paper.}
}

\begin{document}

\title{A Controlled Benchmark of Visual State-Space Backbones with Domain-Shift and Boundary Analysis for Remote-Sensing Segmentation}

\maketitle
\AddToShipoutPictureFG*{%
  \AtPageUpperLeft{%
    \hspace{0.8cm}\raisebox{-1.2cm}{%
      \parbox{0.7\textwidth}{\textit{Accepted for publication at IEEE IGARSS 2026}}%
    }%
  }%
}

\begin{abstract}
Visual state-space models (SSMs) are increasingly promoted as efficient alternatives to Vision Transformers, yet their practical advantages remain unclear under fair comparison because existing studies rarely isolate encoder effects from decoder and training choices. We present a strictly controlled benchmark of representative visual SSM families, including VMamba, MambaVision, and Spatial-Mamba, for remote-sensing semantic segmentation, in which only the encoder varies across experiments. Evaluated on LoveDA and ISPRS Potsdam under a unified 4-stage feature interface and a fixed lightweight decoder, the benchmark reveals three main findings, intra-family scaling yields only modest gains, cross-domain generalization is strongly asymmetric, and boundary delineation is the dominant failure mode under distribution shift. Although visual SSMs achieve favorable accuracy-efficiency trade-offs relative to the controlled CNN and Transformer baselines considered here, the results suggest that future improvements are more likely to come from robustness-oriented design and boundary-aware decoding than from encoder scaling alone. By isolating encoder behavior under a unified and reproducible protocol, this study establishes a practical reference benchmark for the design and evaluation of future Mamba-based segmentation backbones. Code available \href{https://github.com/Dineth14/Mamba-Segmentation}{here}.
\end{abstract}


\section{Introduction} \label{sec:intro} Remote sensing (RS) semantic segmentation must capture fine local structures (e.g., roads and building boundaries) while integrating broad scene context over high-resolution imagery. Convolutional encoder-decoder networks remain strong due to efficient local feature extraction~\cite{aburaed2023semantic,Chen_2018_ECCV}, whereas Vision Transformers (ViTs) improve global context aggregation but incur quadratic attention cost with token count, creating memory and throughput constraints at high resolution~\cite{vaswani2017attention,wang2021unetformer,11217111}. Recently, selective state space models (SSMs) such as Mamba have emerged as an alternative long-range modeling mechanism with more favorable scaling than attention in theory~\cite{gu2023mamba}. Consequently, there is a rapidly growing adoption of SSM-based architectures in remote sensing, driven by their ability to model global context with lower computational overhead than Transformers~\cite{ma2024rsmamba,zhu2024samba,wijenayake2025mamba,wijenayake2025precision}. Vision SSM backbones adapt this idea to 2D feature maps via structured spatial traversal patterns, including VMamba~\cite{liu2024vmamba}, MambaVision~\cite{hatamizadeh2024mambavision}, and Spatial-Mamba~\cite{xiao2025spatial}. Despite growing interest, the practical accuracy--efficiency trade-offs of visual SSM backbones in RS semantic segmentation remain difficult to interpret under fair comparison because reported results often vary across decoder heads, feature interfaces, training schedules, and data pipelines. This is especially important for RS deployments where models are applied tile-wise over large areas and must remain robust under domain shifts. LoveDA explicitly exposes such shifts between Urban and Rural scenes, motivating source only cross domain evaluation under Urban to Rural and Rural to Urban transfer without adaptation~\cite{wang2021loveda}.

To reduce confounding factors, we present a strictly controlled encoder benchmark for visual SSM backbones. We fix the entire segmentation pipeline, data preprocessing and augmentations, input resolution, optimization schedule, loss, evaluation protocol, and a CNN-based lightweight U-Net style decoder~\cite{ronneberger2015unet} with a unified 4-stage feature interface and vary only the encoder backbone. We evaluate on LoveDA~\cite{wang2021loveda} to analyze cross-domain generalization and on ISPRS Potsdam~\cite{rottensteiner2012isprs} to validate conclusions on very high resolution aerial imagery with fine grained objects and boundaries. 

While mean Intersection-over-Union (mIoU) is the standard metric, it provides limited insight into where and why errors occur under domain shift. In high-resolution RS, segmentation failures are often dominated by boundary ambiguity due to mixed pixels, label rasterization, and inconsistent geometric cues across domains. We therefore complement mIoU with boundary focused diagnostics to better characterize failure modes across backbones.

\begin{figure*}[t!]
  \centering
  \vspace{-2mm}
    \includegraphics[width=\textwidth,height=0.22\textheight,keepaspectratio]{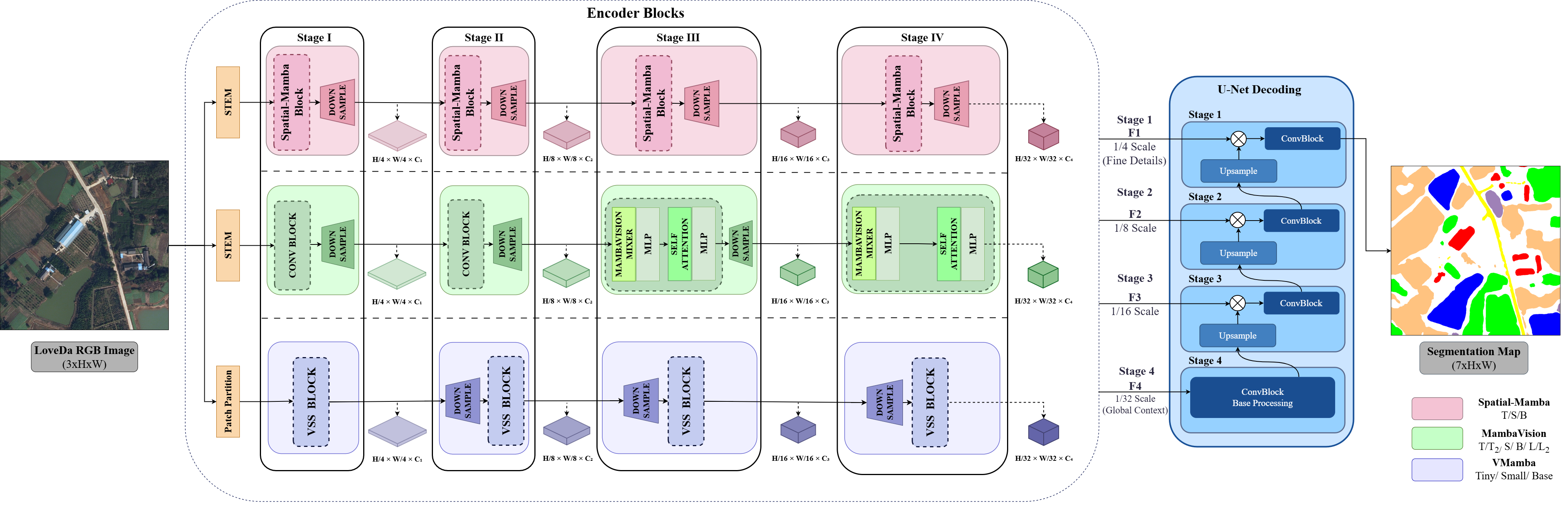}
  \vspace{-2mm}
  \caption{Overview of the controlled benchmark pipeline. To ensure fair comparison, all backbones including visual SSMs (VMamba, MambaVision, Spatial-Mamba) and baselines (CNNs, Transformers) are interfaced through a unified 4-stage feature pyramid and evaluated with the same lightweight U-Net decoder under an identical training and evaluation protocol.}
  \label{fig:arch_overview}
  \vspace{-2mm}
\end{figure*}

Our contributions are three fold, (1) We present a reproducible, strictly controlled benchmark of three visual SSM backbone families (VMamba, MambaVision, Spatial-Mamba) under an identical segmentation pipeline and a unified 4-stage feature interface; (2) We analyze in-domain accuracy,cross-domain robustness, boundary sensitivity, and practical efficiency under a unified measurement protocol and (3) We validate findings on both LoveDA and ISPRS Potsdam to support deployment-oriented model selection under domain shift.

\section{Methodology}
\label{sec:method}

\subsection{Controlled Backbone Benchmark and Visual SSM Backbones}
\label{sec:controlled_backbone}

Selective state space models (SSMs) provide an alternative long-range modeling mechanism with favorable linear scaling compared to quadratic attention mechanisms~\cite{gu2023mamba,vaswani2017attention}. Recent vision backbones adapt this selective recurrence to 2D feature maps via structured spatial scanning and connectivity. We benchmark three representative visual SSM families: VMamba~\cite{liu2024vmamba}, MambaVision~\cite{hatamizadeh2024mambavision}, and Spatial-Mamba~\cite{xiao2025spatial}.
VMamba uses four-directional cross-scan, MambaVision combines Mamba blocks with attention, and Spatial-Mamba applies structure-aware state fusion.

\subsection{Strictly controlled encoder protocol.}
To isolate the impact of the encoder architecture, we evaluate all backbones under an identical segmentation pipeline: the same decoder, loss, training schedule, augmentations, input resolution, and evaluation. For the CNN/Transformer references, we extract only the encoder backbones (e.g., a DeepLabv3-style CNN encoder and a UNetFormer-style Transformer encoder) and plug them into our fixed lightweight U-Net decoder. Consequently, results reflect encoder behavior under a standardized decoder and should not be interpreted as reproducing the original end-to-end DeepLabv3/UNetFormer architectures. All encoders output a unified 4-stage feature pyramid $\{F_1, F_2, F_3, F_4\}$ with output strides $\{4, 8, 16, 32\}$, consumed by the same decoder.

Inputs are processed as $512\times512$ RGB patches with ImageNet normalization.
During training, we apply random crops, horizontal/vertical flips, random $90^\circ$ rotations, color jitter, and Gaussian blur.
We optimize for 50k iterations using AdamW~\cite{loshchilov2017decoupled} with distinct learning rates for the backbone and decoder ($6\times10^{-5}$ and $3\times10^{-4}$, respectively), a weight decay of 0.05, and Poly learning-rate decay ($p=0.9$) with a batch size of 4. A single backbone learning rate isolates representational from optimization effects.

\textbf{Loss.}
We use a three-term objective $L = L_{\text{Lov\'asz}} + L_{\text{Focal}} + 0.5\,L_{\text{Boundary}}$.
We employ Lov\'asz-Softmax loss~\cite{berman2018lovasz} to directly optimize the Jaccard index and Focal loss~\cite{lin2017focal} to address class imbalance. The boundary term $L_{\text{Boundary}}$ penalizes misclassifications within 2 pixels of ground-truth class edges via binary cross-entropy on edge-dilated masks.

\subsection{Unified 4-Stage Feature Interface}
\label{sec:feature_interface}

All encoders expose a 4-scale feature pyramid $\{\mathbf{F}_1,\mathbf{F}_2,\mathbf{F}_3,\mathbf{F}_4\}$ at output strides $\{4,8,16,32\}$ relative to the input image $\mathbf{I}\in\mathbb{R}^{H\times W\times 3}$.
Each feature satisfies $\mathbf{F}_i\in\mathbb{R}^{\frac{H}{s_i}\times \frac{W}{s_i}\times C_i}$ where $s_i\in\{4,8,16,32\}$.
For backbones with matching native strides, we extract stage outputs directly. Otherwise, we add $1\times1$ projection layers to match the expected spatial resolutions and normalize channel dimensions to $\{64,128,256,512\}$ for stages $\{1,2,3,4\}$, respectively.

\begin{table*}[t]
\centering
\begin{threeparttable}
\caption{Controlled benchmark results on LoveDA and ISPRS Potsdam. U$\rightarrow$R and R$\rightarrow$U denote source-only cross-domain mIoU. Top-3 per column: \textcolor{red}{1st}, \textcolor{blue}{2nd}, \textcolor{green!70!black}{3rd}.}
\label{tab:main_merged}

\footnotesize
\setlength{\tabcolsep}{3.0pt}

\begin{tabular}{llcccccccccccc}
\toprule
\textbf{Type} & \textbf{Backbone} &
\textbf{Params} & \textbf{FLOPs} & \textbf{Mem} & \textbf{FPS} &
\multicolumn{5}{c}{\textbf{LoveDA}} &
\multicolumn{3}{c}{\textbf{Potsdam}} \\
& & (M) & (G) & (GB) & (img/s) & mIoU & OA & F1 & U$\rightarrow$R & R$\rightarrow$U & mIoU & OA & F1 \\
\midrule
CNN & DeepLabv3 & 36.0 & 105.3 & 0.40 & 18.52 & 43.01 & 59.14 & 64.20 & 30.36 & 39.98 & 75.09 & 89.06 & 84.44 \\
Transformer & UNetFormer & 16.2 & 48.9 & 0.18 & 40.60 & 48.61 & 65.04 & 63.99 & 34.56 & 44.84 & 74.99 & 89.12 & 84.46 \\
\midrule
SSM & VMamba-Tiny & 27.6 & 26.1 & 0.41 & 18.44
& 55.19 & 72.15 & 70.30 & 35.42 & 53.19
& 77.06 & 90.16 & \textcolor{green!70!black}{86.06} \\
SSM & VMamba-Small & 49.1 & 42.5 & 0.49 & 10.50
& \textcolor{red}{55.66} & \textcolor{red}{73.49} & \textcolor{red}{70.69} & \textcolor{blue}{40.62} & 53.52
& \textcolor{red}{77.59} & \textcolor{blue}{90.43} & \textcolor{red}{86.46} \\
SSM & VMamba-Base & 81.7 & 67.3 & 0.70 & 7.77
& \textcolor{blue}{55.61} & \textcolor{blue}{72.89} & \textcolor{blue}{70.54} & 35.82 & \textcolor{red}{55.50}
& \textcolor{blue}{77.43} & \textcolor{red}{90.61} & \textcolor{blue}{86.19} \\
\midrule
SSM & MambaVision-T & 36.8 & 44.9 & 0.19 & 40.83
& 54.39 & 71.00 & 69.88 & 38.12 & 53.89
& 76.59 & 90.01 & 85.65 \\
SSM & MambaVision-T2 & 40.1 & 50.7 & 0.21 & 34.87
& 54.93 & 72.24 & 69.97 & 38.17 & 51.02
& 76.65 & 90.00 & 85.72 \\
SSM & MambaVision-S & 55.6 & 65.6 & 0.27 & 30.19
& 53.71 & 71.19 & 69.07 & 39.47 & 53.46
& 76.77 & 90.13 & 85.77 \\
SSM & MambaVision-B & 104.2 & 123.8 & 0.45 & 17.72
& 55.17 & 72.42 & \textcolor{green!70!black}{70.43} & \textcolor{green!70!black}{40.24} & 51.96
& \textcolor{green!70!black}{77.08} & 90.27 & 86.03 \\
SSM & MambaVision-L & 236.6 & 273.6 & 0.97 & 8.64
& \textcolor{green!70!black}{55.25} & \textcolor{green!70!black}{72.79} & 70.39 & 38.53 & \textcolor{green!70!black}{54.01}
& 77.07 & \textcolor{green!70!black}{90.30} & 85.98 \\
SSM & MambaVision-L2 & 250.2 & 297.5 & 1.03 & 7.83
& 54.97 & 71.54 & 70.17 & \textcolor{red}{41.74} & \textcolor{blue}{54.62}
& 76.89 & 90.18 & 85.85 \\
\midrule
SSM & Spatial-Mamba-T & 30.7 & 33.3 & 0.21 & 25.76
& 45.56 & 65.31 & 61.85 & 33.10 & 45.30
& 68.23 & 84.78 & 79.46 \\
SSM & Spatial-Mamba-S & 47.1 & 46.7 & 0.27 & 17.64
& 46.54 & 65.57 & 62.88 & 29.72 & 43.71
& 69.21 & 85.48 & 80.18 \\
SSM & Spatial-Mamba-B & 100.5 & 92.3 & 0.52 & 10.97
& 48.03 & 66.35 & 64.19 & 35.23 & 46.55
& 70.00 & 85.92 & 80.81 \\
\bottomrule
\end{tabular}

\begin{tablenotes}
\footnotesize
\item FPS measured at 512$\times$512, batch size 4, FP32 (AMP off), 300 iterations with 50 warmup iterations.
\end{tablenotes}

\end{threeparttable}
\end{table*}

\section{Experimental Setup and Evaluation}
\label{sec:experiments}

All models are trained and evaluated under the strictly controlled protocol in Section~\ref{sec:controlled_backbone}.
We evaluate on LoveDA~\cite{wang2021loveda} to analyze cross-domain generalization and on ISPRS Potsdam~\cite{rottensteiner2012isprs} to assess fine-grained urban scene parsing at very high resolution.

\subsubsection{Datasets and Splits}
LoveDA contains 2,713 Urban images and 3,274 Rural images with 7 land-cover classes at 0.3\,m resolution~\cite{wang2021loveda}.
We use the official train/val splits (Urban: 1,156/321, Rural: 1,366/410) and report:
(i) All to All (train on Urban+Rural, evaluate on the combined validation set), and
(ii) source-only cross-domain generalization under Urban to Rural (U$\rightarrow$R) and Rural to Urban (R$\rightarrow$U), where models are trained on one domain and directly evaluated on the other without adaptation.

ISPRS Potsdam provides 38 tiles at 5\,cm resolution with 6 semantic classes~\cite{rottensteiner2012isprs}.
Following the standard split (24 train tiles / 14 val tiles), we crop each $6000\times6000$ tile into $512\times512$ RGB patches with stride 256 during training. Validation uses non-overlapping stride 512 patches to avoid overlap in evaluation.

\subsubsection{Metrics and Repeats}
We report mean Intersection-over-Union (mIoU) as the primary metric, along with Overall Accuracy (OA) and F1-score, including per-class IoU and F1 derived from the confusion matrix.
Unless stated otherwise, to quantify run-to-run variability, we repeat experiments with three random seeds for the top-performing model in each SSM family and report mean$\pm$std mIoU on LoveDA.
The observed standard deviations are low (VMamba-Small: 0.23, MambaVision-Large: 0.18, Spatial-Mamba-Base: 0.26), indicating stable training under our protocol (Table~\ref{tab:main_merged}).

% Tables/perclass_LOVEDA.tex  (BODY ONLY: no table/table*, no caption/label)
\begin{threeparttable}
\scriptsize
\setlength{\tabcolsep}{2.0pt}
\renewcommand{\arraystretch}{1.08}
\caption{Per-class IoU (\%) on LoveDA All$\rightarrow$All (train on Urban+Rural, evaluate on combined val) for the top model per family.}
\label{tab:per_loveda}

\begin{tabular}{l*{7}{S[table-format=2.2]}}
\toprule
\textbf{Backbone} &
\textbf{Building} &
\textbf{Road} &
\textbf{Water} &
\textbf{Barren} &
\textbf{Forest} &
\textbf{Agriculture} &
\textbf{Background} \\
\midrule
VMamba-Small      & 64.93 & 57.15 & 73.05 & 35.06 & 41.70 & 61.15 & 56.61 \\
MambaVision-L     & 65.66 & 57.18 & 71.04 & 34.85 & 42.22 & 61.33 & 54.47 \\
Spatial-Mamba-B   & 57.54 & 54.08 & 59.72 & 29.67 & 36.02 & 48.79 & 50.41 \\
\bottomrule
\end{tabular}

\end{threeparttable}

\subsubsection{Efficiency Evaluation}
To compare computational cost fairly, we report Params/FLOPs and measure inference throughput (FPS) and peak GPU memory using a unified setup.
All timing and memory numbers are measured at batch size 4 after warm-up on an NVIDIA Quadro GV100 (32\,GB), using PyTorch 2.0.1 and CUDA 11.5, consistent with Section~\ref{sec:controlled_backbone}.

\subsubsection{Pretraining}
All encoder backbones are initialized from ImageNet-1K pretrained weights.

\begin{figure*}
  \centering
  \begin{minipage}[t]{0.49\textwidth}
    \centering
    \includegraphics[width=\linewidth,height=0.33\textheight,keepaspectratio]{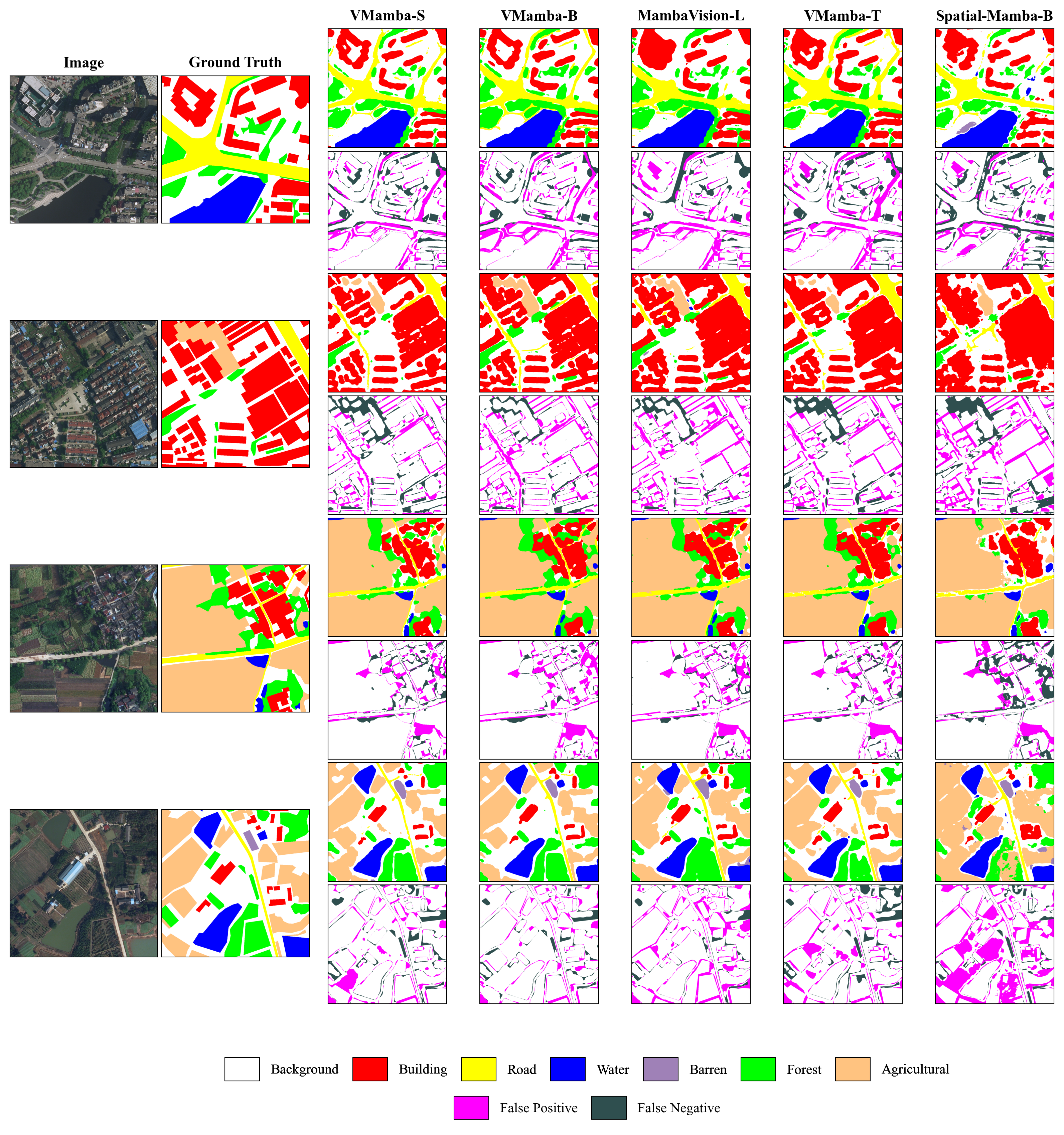}
    \vspace{-1mm}
  \end{minipage}
  \hfill
  \begin{minipage}[t]{0.49\textwidth}
    \centering
    \includegraphics[width=\linewidth,height=0.33\textheight,keepaspectratio]{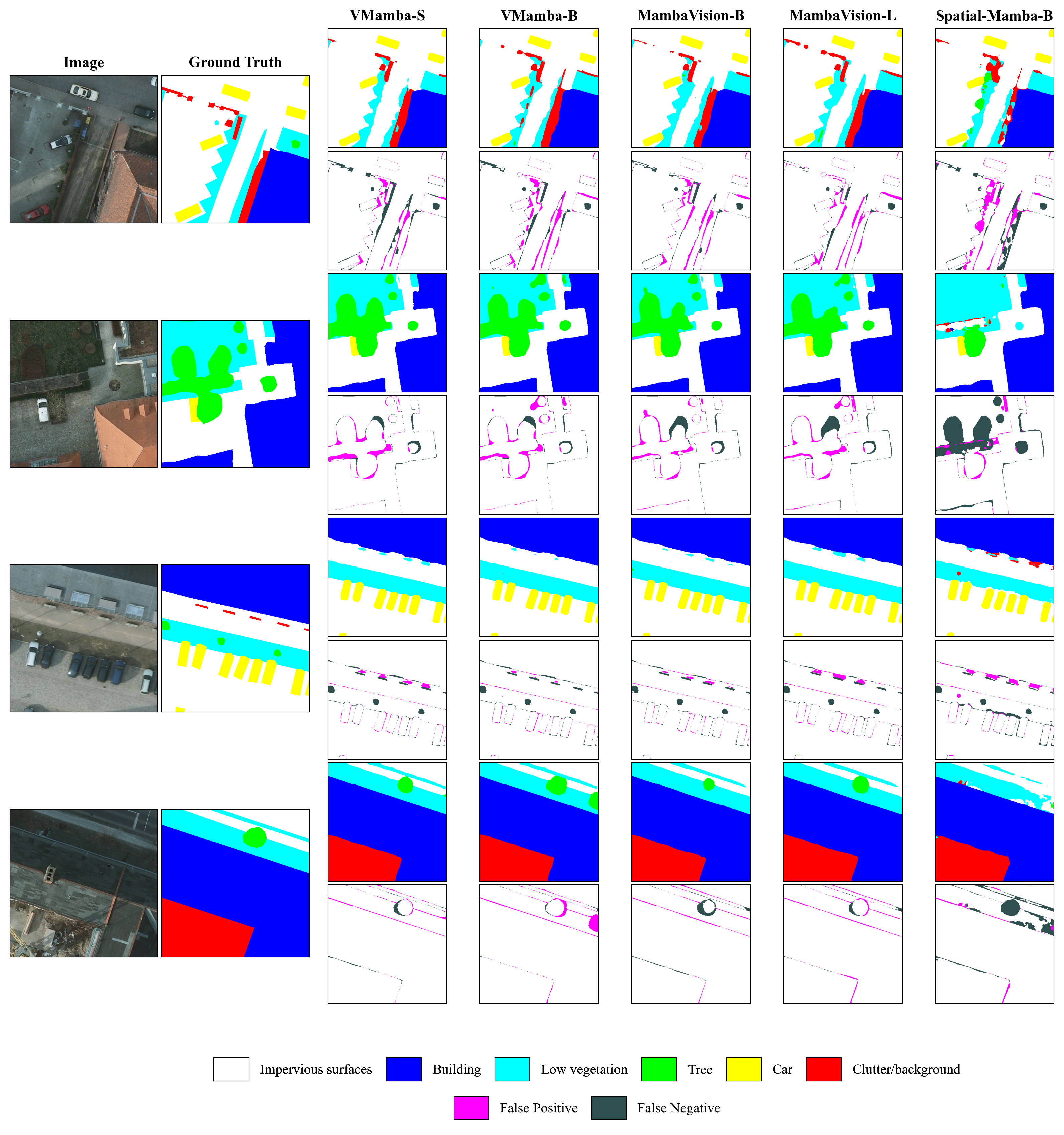}
    \vspace{-1mm}
  \end{minipage}
  \vspace{-1mm}
  \caption{Qualitative comparison under the controlled protocol. (a) LoveDA: Urban/Rural examples illustrating boundary delineation and robustness under domain shift. (b) ISPRS Potsdam: fine-grained structures (buildings, impervious surfaces, vegetation, cars) in very-high-resolution aerial imagery.}
  \label{fig:qualitative_side_by_side}
  \vspace{-2mm}
\end{figure*}

\section{Results and Discussion}
\label{sec:discussion}

\subsection{Overall benchmark trends}
\begin{figure}
  \centering
  % Make sure the image file is named 'fps_vs_miou.png' or update the name below
  \includegraphics[width=0.95\linewidth]{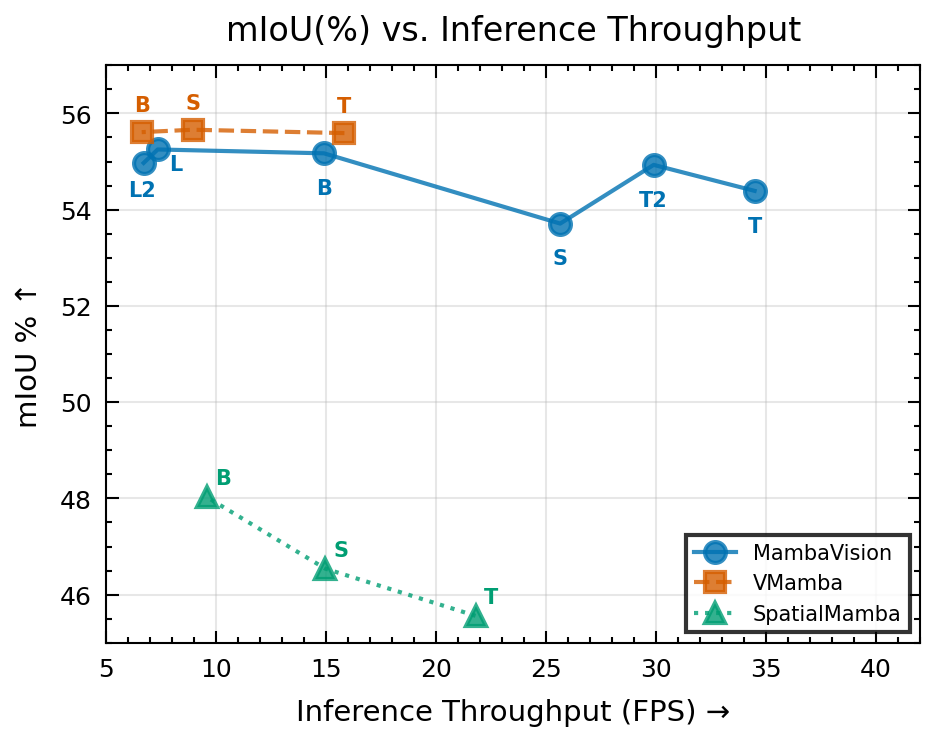} 
  \caption{Efficiency analysis on LoveDA: A comparison of mIoU versus FPS across VSS backbones.}
  \label{fig:efficiency_tradeoff}
\end{figure}

Among higher-capacity models, MambaVision-L2~\cite{hatamizadeh2024mambavision} yields the best Urban to Rural generalization (\textbf{41.74} mIoU) but incurs high cost (250.2M params). Notably, VMamba-Small remains competitive (40.62 mIoU) at a fraction of the complexity (Fig.~\ref{fig:efficiency_tradeoff}). Conversely, Spatial-Mamba~\cite{xiao2025spatial} underperforms on LoveDA (45.56--48.03 mIoU), suggesting that for fragmented remote-sensing scenes, feature structure and decoder-friendliness outweigh initialization.

\subsection{Per-class behavior on LoveDA and Potsdam}
The breakdowns Per-class (Table~\ref{tab:per_loveda},~\ref{tab:perclass_potsdam})reveal failure modes resistant to encoder scaling. On LoveDA, Barren is the most difficult category (35.06 IoU for VMamba-Small) due to texture ambiguity, while on Potsdam, Clutter/Background consistently scores lowest. These trends imply that future improvements must address boundary-aware decoding and ambiguity handling rather than relying solely on larger backbones.

% Tables/perclass_POSTDAM.tex  (BODY ONLY: no table/table*, no caption/label)
\begin{threeparttable}
\scriptsize
\setlength{\tabcolsep}{3.0pt}
\renewcommand{\arraystretch}{1.08}
\caption{Per-class IoU on ISPRS Potsdam.}
\label{tab:perclass_potsdam}
\begin{tabular}{l*{6}{S[table-format=2.2]}}
\toprule
\textbf{Backbone} &
\multicolumn{1}{c}{\textbf{\shortstack{Impervious\\Surface}}} &
\multicolumn{1}{c}{\textbf{Building}} &
\multicolumn{1}{c}{\textbf{\shortstack{Low\\Vegetation}}} &
\multicolumn{1}{c}{\textbf{Tree}} &
\multicolumn{1}{c}{\textbf{Car}} &
\multicolumn{1}{c}{\textbf{\shortstack{Clutter/\\Background}}} \\
\midrule
VMamba-Small    & 85.33 & 93.18 & 76.74 & 78.19 & 85.98 & 46.15 \\
MambaVision-B   & 84.98 & 93.57 & 76.29 & 77.54 & 85.93 & 44.15 \\
Spatial-Mamba-B & 78.72 & 88.58 & 69.99 & 68.51 & 81.28 & 32.90 \\
\bottomrule
\end{tabular}
\end{threeparttable}

\subsection{Diagnostic analysis, boundary vs.\ interior}
To identify what limits cross-domain robustness, we evaluate boundary and interior mIoU using a $\tau{=}2$ pixel dilation protocol (Table~\ref{tab:diagnostic_analysis}). Boundary pixels are Sobel edges dilated by a 3$\times$3 element. The radius matches Section~\ref{sec:controlled_backbone}.
A consistent pattern emerges: for the strongest backbones, boundary accuracy is substantially lower than interior accuracy.
For example, VMamba-Tiny/Small achieve boundary mIoU of \textbf{28--29} while interior mIoU reaches \textbf{58--59}, 
producing a gap of roughly \textbf{30} mIoU (Table~\ref{tab:diagnostic_analysis}, $\Delta = \text{mIoU}_{\text{interior}} - \text{mIoU}_{\text{boundary}}$).
This indicates that even stronger encoders struggles to resolve geometric ambiguity at edges (thin roads, building outlines, narrow water boundaries),
making boundary delineation a primary bottleneck.

Importantly, the boundary deficit is not uniform across backbone families, weaker boundary/interior scores for Spatial-Mamba variants (Table~IV) align with their lower overall mIoU (Table~I),
supporting the conclusion that cross-domain failures concentrate on boundary-heavy pixels rather than on interior regions alone.

\begin{table}[t]
\centering
\caption{Diagnostic analysis on LoveDA. Boundary/Interior mIoU use $\tau=2$ pixel dilation (see Section IV-A). Class groups: Structural = \{Building, Road\}, Natural = \{Forest, Agriculture, Water\}.}
\label{tab:diagnostic_analysis}
\setlength{\tabcolsep}{2pt}
\renewcommand{\arraystretch}{1.1}
\resizebox{\linewidth}{!}{%
\begin{tabular}{l|ccc|cccc}
\hline
\textbf{Backbone} & \textbf{Boundary} & \textbf{Interior} & $\boldsymbol{\Delta}$ &
\textbf{R$\rightarrow$U Struct} & \textbf{R$\rightarrow$U Nat} &
\textbf{U$\rightarrow$R Struct} & \textbf{U$\rightarrow$R Nat} \\
\hline
VMamba-Tiny   & 29.0 & 58.5 & 29.5 & 55.9 & 57.7 & 41.4 & 34.2 \\
VMamba-Small  & 28.2 & 59.1 & 30.8 & 56.0 & 58.2 & 42.3 & 46.2 \\
VMamba-Base   & 29.4 & 58.9 & 29.5 & 58.1 & 63.0 & 46.4 & 40.9 \\
\hline
MambaVision-T & 29.5 & 57.6 & 28.0 & 56.8 & 55.0 & 39.3 & 41.1 \\
MambaVision-S & 27.0 & 57.0 & 30.0 & 55.7 & 60.5 & 39.2 & 47.1 \\
MambaVision-B & 29.0 & 58.4 & 29.4 & 57.0 & 54.9 & 44.5 & 42.7 \\
\hline
SpatialM-Tiny & 17.9 & 28.4 & 10.5 & 44.9 & 50.4 & 31.4 & 35.3 \\
SpatialM-Small& 19.4 & 33.9 & 14.5 & 42.9 & 48.0 & 30.8 & 29.6 \\
SpatialM-Base & 26.6 & 50.8 & 24.2 & 43.9 & 49.5 & 29.9 & 32.0 \\
\hline
\end{tabular}%
}
\vspace{-2mm}
\end{table}

\subsection{Cross-domain behavior and asymmetry}
Table~I shows a consistent asymmetry: Rural to Urban outperforms Urban to Rural across backbones (e.g., VMamba-Small: 40.62$\rightarrow$53.52, MambaVision-L2: 41.74$\rightarrow$54.62). We attribute this to the greater appearance diversity of rural scenes, which promotes transferable representations, whereas urban-trained models tend to overfit to regularized geometry and sharp edges that do not hold in rural imagery. Table~\ref{tab:diagnostic_analysis} further indicates Urban to Rural degradation is dominated by \emph{natural} categories (forest/agriculture/water), where texture and context shifts increase boundary ambiguity. Overall, these results point to decoder-side boundary/uncertainty modeling as the most promising route to improve cross-domain robustness~\cite{aburaed2023semantic,wang2021loveda}. Differing urban/rural edge types and SSM fixed scan paths amplify this deficit under shift.

\section{Conclusion}
\label{sec:Conclusion}

We presented a controlled benchmark of visual SSM backbones for remote-sensing semantic segmentation, evaluating VMamba~\cite{liu2024vmamba}, MambaVision~\cite{hatamizadeh2024mambavision}, and Spatial-Mamba~\cite{xiao2025spatial} against CNN (DeepLabv3~\cite{Chen_2018_ECCV}) and Transformer (UNetFormer~\cite{wang2021unetformer}) baselines on LoveDA~\cite{wang2021loveda} and ISPRS Potsdam~\cite{rottensteiner2012isprs} under an identical decoder and training protocol. Overall, SSM encoders provide the most favorable accuracy--efficiency trade-off and are consistently competitive to the CNN/Transformer baselines under the controlled setting. We also observe a clear cross-domain asymmetry, where transferring from rural to urban scenes is systematically easier than the reverse. Finally, diagnostic evaluation highlights boundary delineation as the dominant source of errors under domain shift, indicating that future improvements should prioritize decoder-side boundary refinement and stronger multi-scale fusion rather than further encoder scaling alone. Broader baseline coverage (e.g., SegFormer, Swin-UNet) and alternative decoder evaluation remain open directions.

\small
\bibliographystyle{IEEEtranN}
\newpage
\bibliography{references}

\end{document}